\documentclass[letterpaper, 10 pt, conference]{ieeeconf}
\usepackage{amsmath,amsfonts}
\usepackage{algorithmic}
\usepackage{algorithm}
\usepackage{array}
\usepackage{subcaption}
\usepackage{textcomp}
\usepackage{stfloats}
\usepackage{url}
\usepackage{verbatim}
\usepackage{graphicx}
\usepackage{cite}
\usepackage{xcolor}
\usepackage{lipsum}
\usepackage{multirow}
\usepackage{tabularx}
\usepackage{float}
\usepackage{booktabs}
\usepackage{bm} 
\usepackage{tikz}
\usepackage{pgfplots}
\usetikzlibrary{3d, positioning, decorations.markings, calc}
\usetikzlibrary {decorations.text}
\usetikzlibrary{angles, quotes}

\pgfplotsset{compat=1.18}

\IEEEoverridecommandlockouts
\title{\LARGE \bf
Dual LiDAR-Based Traffic Movement Count Estimation at a Signalized Intersection: Deployment, Data Collection, and Preliminary Analysis
}

\author{
    Saswat~Priyadarshi~Nayak$^{1}$,~Guoyuan~Wu$^{2}$,~Kanok~Boriboonsomsin$^{3}$,~Matthew~Barth$^{4}$%
    \thanks{Saswat Priyadarshi Nayak (Corresponding author, snaya004@ucr.edu), Guoyuan Wu, Matthew Barth are with the Department of Electrical and Computer Engineering, University of California Riverside, Riverside, USA.}
    \thanks{Kanok Boriboonsomsin is with the Center for Environmental Research and Technology (CE-CERT), University of California Riverside, Riverside, USA.}
}

\begin{document}

\maketitle
\thispagestyle{empty}
\pagestyle{empty}

\begin{abstract}

Traffic Movement Count (TMC) at intersections is crucial for optimizing signal timings, assessing the performance of existing traffic control measures, and proposing efficient lane configurations to minimize delays, reduce congestion, and promote safety. Traditionally, methods such as manual counting, loop detectors, pneumatic road tubes, and camera-based recognition have been used for TMC estimation. Although generally reliable, camera-based TMC estimation is prone to inaccuracies under poor lighting conditions during harsh weather and nighttime. In contrast, Light Detection and Ranging (LiDAR) technology is gaining popularity in recent times due to reduced costs and its expanding use in 3D object detection, tracking, and related applications. This paper presents the authors' endeavor to develop, deploy and evaluate a dual-LiDAR system at an intersection in the city of Rialto, California, for TMC estimation. The 3D bounding box detections from the two LiDARs are used to classify vehicle counts based on traffic directions, vehicle movements, and vehicle classes. This work discusses the estimated TMC results and provides insights into the observed trends and irregularities. Potential improvements are also discussed that could enhance not only TMC estimation, but also trajectory forecasting and intent prediction at intersections.
\end{abstract}

\section{INTRODUCTION}
\label{sec: introduction}

Over the past few decades, the resulting surge in travel demand has contributed to increased congestion, safety risks, and environmental concerns~\cite{Talebian2018PredictingInnovations}. It is essential to develop smart traffic monitoring systems and design more efficient road and traffic control infrastructure to better address these challenges. Traffic Movement Count (TMC) at an intersection involves keeping a count of different classes of vehicles performing maneuvers involving thru, left-turn, right-turn, and U-turn movements. TMC estimation helps to understand traffic patterns based on different travel directions. Also, it helps to identify the root cause of delays and congestion during both peak and off-peak hours~\cite{article}. 

TMC estimation technologies can be divided into two categories - intrusive and non-intrusive. Intrusive technologies require physical installation on or under the road surfaces. These include well understood and tested methods such as inductive loop detectors, pneumatic road tubes, piezoelectric sensors, and magnetometers~\cite{McGowen2011AccuracyOP, article2, 928860}. These sensors are calibrated and offer reliable vehicle detections. However, they are subject to wear and tear over time and offer limited flexibility for temporary traffic studies that require different configurations.

On the other hand, non-intrusive technologies involve the use of cameras, radars, acoustic sensors, infrared, drones, and in recent days, Light Detection and Ranging (LiDAR) sensors~\cite{Aljamal2021Real-TimeData}. Among these, camera-based object detection, tracking, and counting (DTC) methods have been quite popular since the last two decades because of low cost and ease of implementation~\cite{Premaratne2023ComprehensiveHighways}. 
Adl et al. proposed a zone-based and path-based vehicle counting algorithm using a fisheye camera in \cite{Adl2024EnhancedSystem}. Although, aforementioned camera-based techniques provided an average accuracy of 85-90\% in TMC estimation, they perform poorly under harsh weather and poor lighting conditions. LiDARs can solve these issues, as they are not only resilient to harsh weather and bad lighting conditions but also highly accurate in their ranging capabilities. Over the last few years, LiDARs have been extensively used in smart infrastructure-based sensing to carry out tasks ranging from 3D detection, tracking, and prediction. LiDAR-based localization techniques~\cite{Bai2022CyberLiDAR, Nayak2024Infrastructure-AssistedMeasurements} offer precise positioning accuracy, making it a suitable candidate for TMC estimation.

This paper describes a project that utilizes two LiDARs mounted on the diagonal traffic poles at a signalized intersection to estimate TMC. The raw point cloud data from the LiDARs are fed to a PointPillars-based 3D detection pipeline~\cite{lang2019pointpillarsfastencodersobject}. The pipeline outputs 3D bounding boxes of the detected road agents, which are then fed to a zone-based TMC estimation module to output vehicle movement counts. The structure of the paper is as follows. Section~\ref{sec: background} discusses related prior work. Section~\ref{sec: dual-lidar setup} presents the dual-LiDAR setup at the intersection. Section \ref{sec: methodology} provides the methodology. The results and potential improvements are discussed in Sections \ref{sec: results_discussion} and \ref{sec: lessons_learned}. Finally, Section \ref{sec: conclusion} concludes the paper.

\section{RELATED WORK}
\label{sec: background}

Traffic monitoring systems involving in-situ traffic detectors, vehicle sensor networks, or probe vehicles have been extensively studied in the literature~\cite{Jain2019ATechniques}. Although LiDAR-based 3D object detection is generally reliable and accurate, its application in traffic movement analysis remains limited. Most LiDAR-related research in Intelligent Transportation Systems (ITS) has focused on advancing cooperative and automated vehicle technologies~\cite{Wolcott2017RobustDriving:}. Leveraging LiDAR for estimating traffic movement counts could enhance the performance of various routing and environmental applications, such as Eco-Routing and Eco-Approach and Departure (EAD).

Jagirdar et al.\cite{Jagirdar2019DevelopmentResults} conducted a proof-of-concept (POC) experiment to evaluate traffic counts using a single 2D LiDAR at intersections. Continuous Wavelet Transformation (CWT) and a Support Vector Machine (SVM) algorithm were used to analyze vehicle distance profiles, yielding a reported vehicle count accuracy of 83–94\%. In a follow-up study\cite{Jagirdar2023DevelopmentSensors}, the authors deployed four two-channel LiDARs at an intersection for turning movement counts. K-means clustering and an inverse sensor model were employed in this work, resulting in an accuracy of 83.8\%. Xie and Rajamani~\cite{Xie2022VehicleLidar} proposed an SVM-based approach for vehicle counting and maneuver classification using a single low-density 2D flash LiDAR, reporting an error rate of 17.54\%.

Wu et al. proposed a video and LiDAR system~\cite{Wu2023ASafety} to analyze intersection safety by tracking pedestrians and vehicles. Although the focus was on detecting conflict points and potential collisions, the system could be extended to traffic count studies. In~\cite{Guan2023EvaluationData}, three roadside cameras and one LiDAR were used to extract trajectory data and evaluate traffic movement counts. It was found that LiDAR-based detection outperformed camera-based detection by 2–3\%. Carranza et al.~\cite{Saldivar-Carranza2024VehicleData} presented traffic signal performance measures using LiDAR-generated trajectory data for vehicles and pedestrians. The trajectory data were used to generate Purdue Probe Diagrams (PPDs), which were then used to estimate traffic movement counts, downstream blockage, level of service, and other performance metrics.

The aforementioned research on LiDAR-based traffic monitoring has produced reliable results but also reveals certain limitations and gaps that need to be addressed. Studies such as~\cite{Jagirdar2019DevelopmentResults, Jagirdar2023DevelopmentSensors, Xie2022VehicleLidar} use 2D LiDAR and statistical measures to evaluate vehicle counts, but are limited to a single lane or traffic direction. The work in~\cite{Guan2023EvaluationData} utilizes a 32-beam LiDAR, which results in limited range and detectability. While~\cite{Saldivar-Carranza2024VehicleData} provides a comprehensive report on the use of LiDAR-generated trajectory data to assess traffic performance measures, it lacks information on vehicle classes and types.

To the best of the authors' knowledge, the work presented in this paper is novel and is the first to address the identified research gaps through the following contributions:

\begin{itemize}
\item Deployment of two high-density Ouster OS1-128 LiDARs at opposite corners of the intersection to maximize coverage.
\item Implementation of a deep learning-based PointPillars model to detect vehicles at the intersection with precise positioning accuracy.
\item Application of a zone-based estimation technique to compute Traffic Movement Count (TMC) across various vehicle classes, travel directions, and movement types.
\item Analysis of the TMC results to identify trends and irregularities, and to propose potential improvements for future work.
\end{itemize}

\section{DUAL-LIDAR DEPLOYMENT}
\label{sec: dual-lidar setup}

\begin{figure}[ht]
    \centering
    \includegraphics[width=0.85\linewidth]{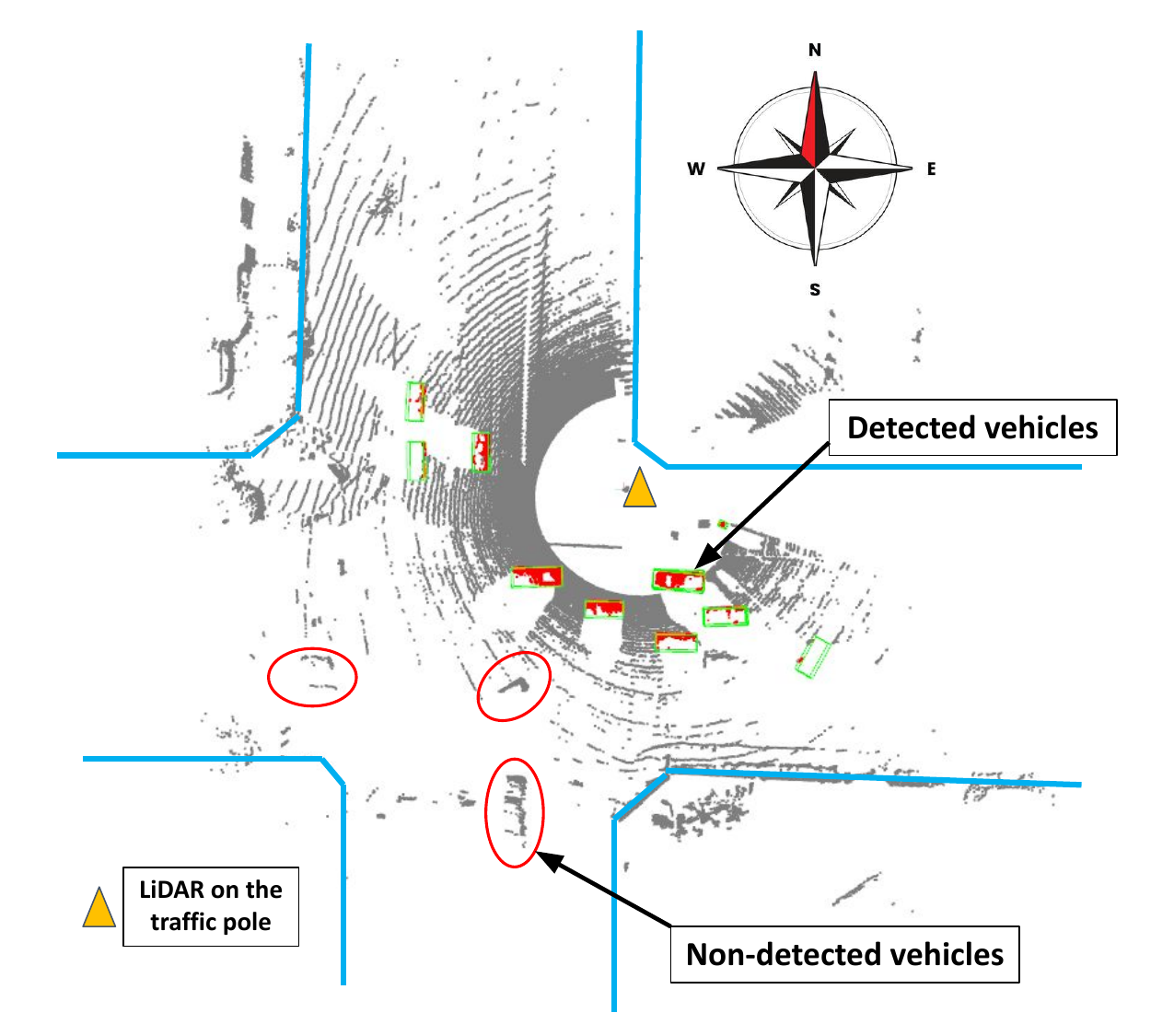}
    \caption{Detection results from a single LiDAR located at the North-East corner of the intersection in Rialto.}
    \label{fig: detection_LIDAR_NE}
\end{figure}

 The motivation for using two LiDARs at the intersection was to cover the entire intersection and provide reliable detection of road agents. Based on initial LiDAR data collection from only one LiDAR at the North-East corner of the intersection, it was concluded that the single LiDAR cannot provide coverage for the whole intersection, as shown in Fig.~\ref{fig: detection_LIDAR_NE}.

A configuration with LiDARs at the North-East and South-West corners would have provided good coverage of large vehicle movements along the southern and western approaches. However, the required Ethernet connection length of approximately 500 ft exceeded the IEEE limit of 328 ft, leading to transmission issues and data sparsity, making the setup impractical.

\begin{figure}[!ht]
    \centering
    \begin{subfigure}{\linewidth}
        \centering
        \includegraphics[width=\linewidth]{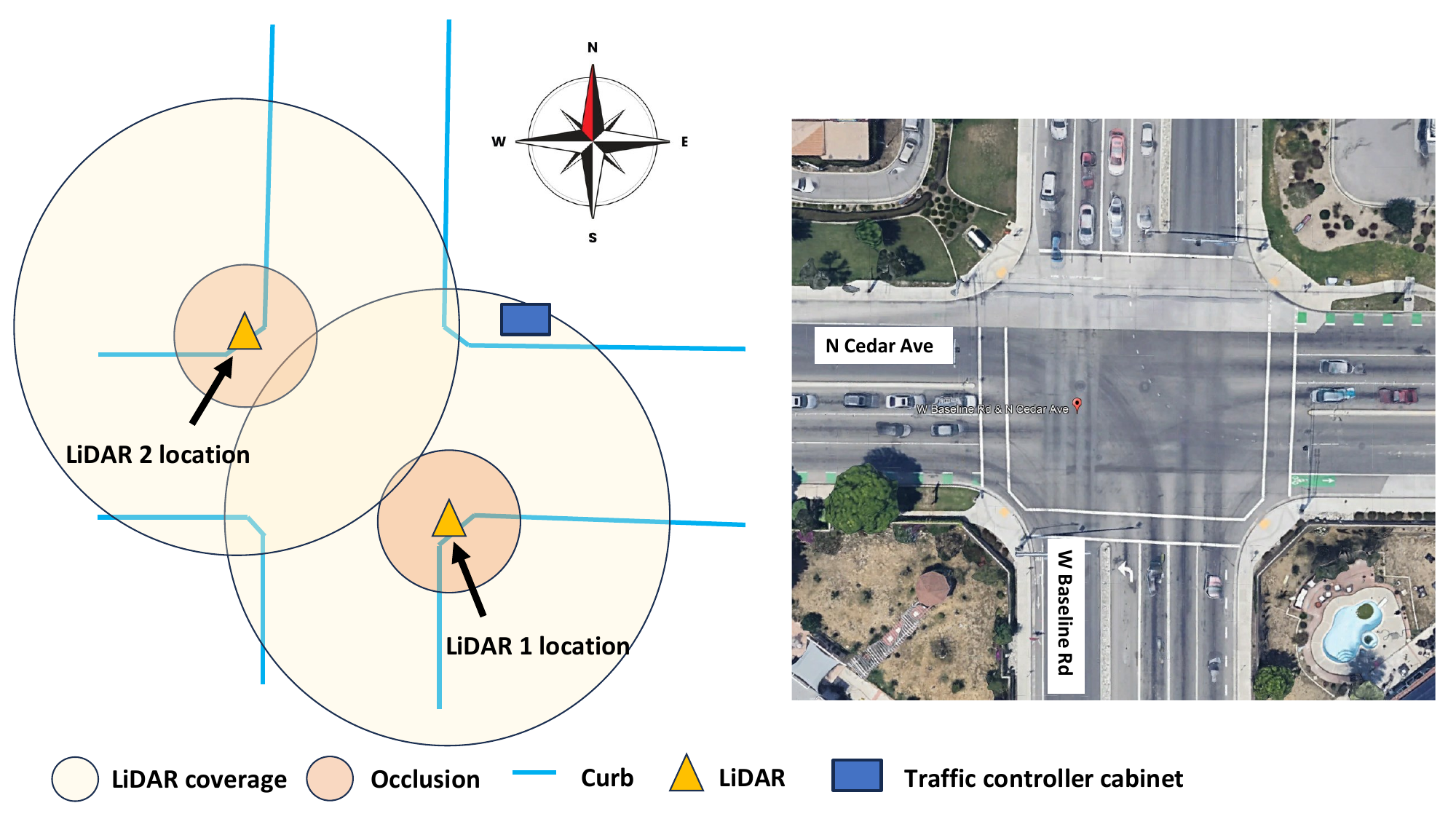}
        \caption{Placement of two LiDARs at the Northwest and Southeast corners of the intersection.}
        \label{fig:dual_lidar_config}
    \end{subfigure}
    
    \vspace{0.5em} 

    \begin{subfigure}{\linewidth}
        \centering
        \includegraphics[width=\linewidth]{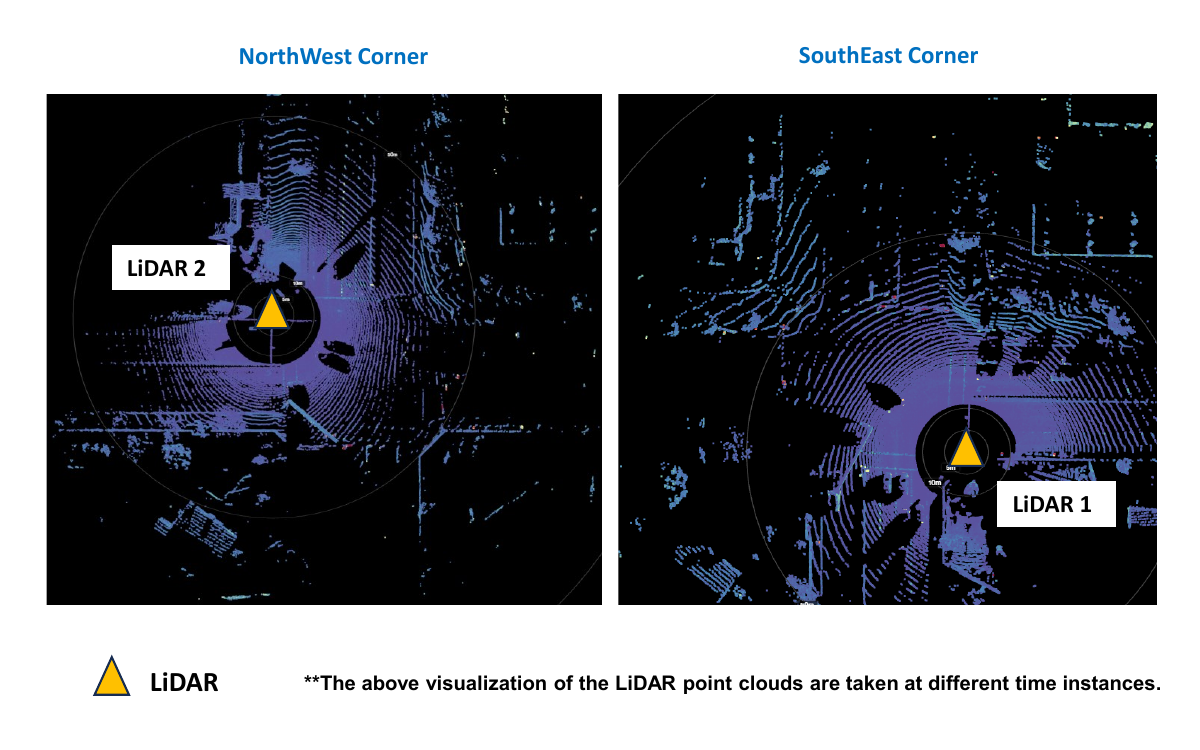}
        \caption{Point cloud visualization from the two LiDARs at the intersection.}
        \label{fig:dual_lidar_visuals}
    \end{subfigure}
    
    \caption{LiDAR deployment configuration and corresponding point cloud visualization.}
    \label{fig:dual_lidar_combined}
\end{figure}

Hence, the LiDARs needed to be brought as close as possible to the traffic cabinet housing the edge computer to ensure the connection lengths to the LiDARs were well within the 328 ft mark. To avoid this issue and ensure overall coverage of the intersection, it was decided that the two LiDARs would be installed at the Northwest and Southeast corners, respectively, as shown in Fig.~\ref{fig:dual_lidar_config}.

\begin{figure}[ht]
    \centering
    \includegraphics[width=\linewidth]{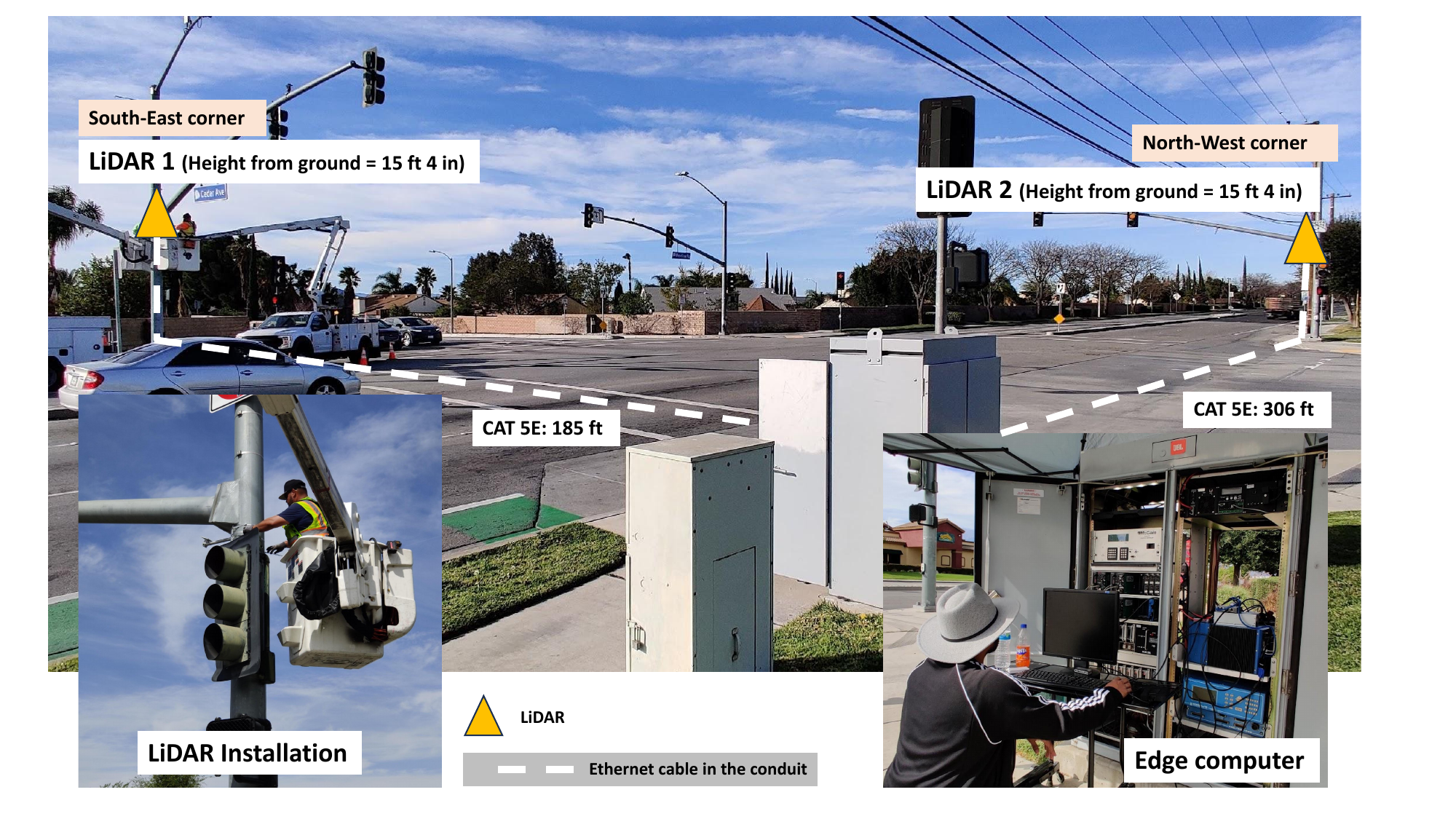}
    \caption{Dual-LiDAR deployment at Baseline and Cedar in the city of Rialto.}
    \label{fig: lidar_deployment}
\end{figure}

The two LiDARs (Ouster OS1-128) were installed on the North-West and South-East corners of the intersection at a height of 15 ft 4 in each and a connection length of 306 ft and 185 ft respectively, as shown in Fig.~\ref{fig: lidar_deployment}. CAT 5E Ethernet cables provided by the installers were used for this deployment. The traffic controller cabinet at the North-East corner of the intersection housed the edge-computer, Ethernet switch, and Power Over Ethernet (POE) injectors. POE injectors are used to supply power to the two LiDARs and simultaneously retrieve data from them through a single Ethernet cable. Furthermore, a POE splitter is used on the LiDAR's side to split the power and data interface for the LiDAR. The point cloud data from the two LiDARs were visualized on the Ouster Visualization software for checking the quality of the data, as shown in Fig.~\ref{fig:dual_lidar_visuals}.

\section{METHODOLOGY}
\label{sec: methodology}

The raw point cloud data coming from the LiDARs was fed to a PointPillars-based detection pipeline running on the edge computer. The detection results (.json format), consisting of detected road agent information (position, bounding box dimensions, direction), were logged as text files (.txt). The raw point cloud data from the LiDARs was also stored as .bin files for each processed LiDAR frame. A validation dataset was collected between 1:51 pm and 2:10 pm (local time) on 20th March, 2024, to assess the accuracy of LiDAR-based detection, tracking, and TMC estimation processes. The dataset contains the following information:

\begin{itemize}
    \item Inference and raw data from the two LiDARs (3-5 Hz)
    \item Drone footage for visual reference (30 FPS)
\end{itemize}

In this study, only the inference results, i.e., bounding box detections from the two LiDARs, and the drone video are used for evaluating the TMC at the intersection. The former is used for TMC estimation, whereas the latter is used to establish the ground truth for TMC. The collected raw point cloud data is not used in the present study but will be utilized in future work for object labeling, enabling detection and tracking performance evaluation.

\begin{figure}[!htb]
    \centering
    \includegraphics[width=\linewidth]{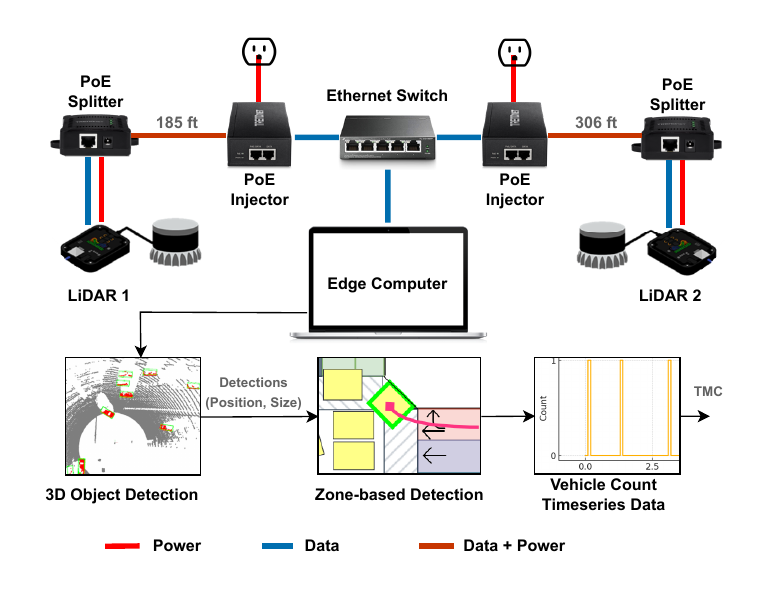}
    \caption{Architecture for Dual-LiDAR operation.}
    \label{fig: dual_lidar_POE}
\end{figure} 

    \subsection{3D Object Detection}

    The 3D point cloud data generated by the LiDARs is processed through a bounding box detection pipeline, which outputs the 3D position (x,y,z), size (length, width, height), and heading angle of detected objects. All detections are represented in the LiDAR coordinate frame and are time-stamped with the corresponding UNIX time.

    For the detection pipeline, we employ the PointPillars~\cite{lang2019pointpillarsfastencodersobject} model from the open-source MMDetection3D framework. Prior studies and experimental evaluations have demonstrated the effectiveness of PointPillars for detecting traffic road agents at intersections using roadside LiDARs~\cite{Bai2022CyberLiDAR}. In this study, only the detected objects' positions and lengths are utilized for TMC estimation.

    \subsection{Coordinate Systems and Transformations}

    The point cloud data and the detections from both the LiDARs are represented in their respective sensor coordinate frames denoted by $\bm{L_1}$ and $\bm{L_2}$, respectively. The global Earth-Centered, Earth-Fixed (ECEF) coordinate system is denoted by $\bm{E}$. A local North East Down (NED) coordinate system denoted by $\bm{N}$ is constructed by choosing an origin point at the intersection, and acts as a common coordinate frame to represent detections and/or tracking results from both LiDARs.

    The transformation between the LiDAR frame and the ECEF frame is given by:

    \begin{equation}
        \begin{bmatrix} \mathbf{P}_{E} & 1 \end{bmatrix} =  
        \begin{bmatrix} \mathbf{P}_{L} & 1 \end{bmatrix}\mathbf{T_{EL}}^T,
    \end{equation}

   where \( \mathbf{P}_{E} \in \mathbb{R}^{3} \) represents the point in the ECEF coordinate frame, 
\( \mathbf{P}_{L} \in \mathbb{R}^{3} \) represents the point in the LiDAR coordinate frame, and 
\( \mathbf{T}_{EL} \in \mathbb{R}^{4 \times 4} \) is the homogeneous transformation matrix that transforms a point from the LiDAR frame to the ECEF frame (\( L \rightarrow E \)). The transformation matrix \( \mathbf{T}_{EL} \) is computed using the Ground Control Point (GCP) georeferencing technique~\cite{Nayak2023EvaluationModels}. The transformation of points from the ECEF coordinate frame to the NED coordinate frame is given by,
    \begin{equation}
        \mathbf{P}_{N} = 
        \mathbf{R_{NE}} 
        (\mathbf{P}_{E} - \mathbf{P}_{E, ref}), 
    \end{equation}

    where \(  \mathbf{P}_{E, ref} \in \mathbb{R}^{3} \) represents the origin of NED coordinate frame in the ECEF coordinate system, \( \mathbf{R}_{NE} \in \mathbb{R}^{3 \times 3} \)  represents the rotation matrix from the ECEF frame to the NED frame, and \( \mathbf{P}_{N} \in \mathbb{R}^{3} \) and \( \mathbf{P}_{E} \in \mathbb{R}^{3} \) represent the position in the NED and ECEF frames, respectively.

    \definecolor{mygreen}{RGB}{0,180,0} 

    \begin{figure}[ht]
        \centering
    \begin{tikzpicture}[scale=2, >=latex, line join=bevel, line cap=round]

        \draw[gray] (0,0) circle (1); 
    
        \draw[thick,->,red] (0,0) -- (-0.5,-0.5) coordinate (A) node[below] {\textbf{$X_{E}$}};
        \draw[thick,->,mygreen] (0,0) -- (1.2,0) node[below] {\textbf{$Y_{E}$}};
        \draw[thick,->,blue] (0,0) -- (0,1.2) node[left] {\textbf{$Z_{E}$}};        

        \draw[dashed, black] (-1,0) arc (180:360:1 and 0.4); 
        \draw[dashed, gray!50] (1,0) arc (0:180:1 and 0.4); 

        \draw[dashed, gray!50] (0,-1) arc (-90:90:0.4 and 1);  
        \draw[dashed, black] (0,1) arc (90:270:0.4 and 1);  

        \path[decorate,decoration={text along path, text align=center, text={|\scriptsize\bfseries|Equator}}]
        (-1,0) arc (180:250:1 and 0.35);

        \path[decorate,decoration={text along path, text align=center, text={|\scriptsize\bfseries|Prime Meridian}}]
        (0,1) arc (94:230:0.4 and 1);

        \begin{scope}[shift={(0.5,0.5)}, rotate=116]
            \draw[thick, green] (-0.2,-0.1) -- (0.2,-0.1) -- (0.3,0.2) -- (-0.1,0.2) -- cycle;
            \fill[green!30, opacity=0.3] (-0.2,-0.1) -- (0.2,-0.1) -- (0.3,0.2) -- (-0.1,0.2) -- cycle;
        \end{scope}

        \draw[thick, orange] (0.45,0.5) arc (60:0:0.2 and 0.96) coordinate (B) -- (0,0) coordinate (O); 

        \begin{scope}[shift={(0.45,0.5)}, rotate=30]
            \draw[thick,->,mygreen] (0,0) -- (0.5,-0.2) node[below, xshift = 5mm] {\scriptsize \textbf{$Y_N (East)$}};
            \draw[thick,->,red] (0,0) -- (0,0.5) node[above, xshift = 5mm] {\scriptsize \textbf{$X_N (North)$}};
            \draw[thick,->,blue] (0,0) -- (-0.4,-0.12) node[ right] {\scriptsize \textbf{$Z_N (Down)$}};
            
        \end{scope}

        \draw[thick, gray] (0,0) -- (0.17,0.20) coordinate (D);  

        \pic [draw, angle radius=0.3cm, angle eccentricity=1.45, "$\lambda$"] {angle=A--O--B};
        \pic [draw, angle radius=0.45cm, angle eccentricity=1.45, "$\varphi$", text height=4.5ex, text depth=0.5ex] {angle=B--O--D};

        \node[draw, fill=white, align=center, text width=2.5cm, minimum height=0.8cm, rounded corners=2pt, baseline] 
          at (2, 1.3) (chatbox)  
          {
            \begin{tikzpicture}[scale=0.7]
                \begin{scope}[shift={(-1cm,-1cm)}] 
                    \draw[thick,->,mygreen] (-2,-1) -- (1,-1) node[above, yshift = -2 mm] {\scriptsize \textbf{$Y_N$}};
                    \draw[thick,->,red] (-2,-1) -- (-2,2) node[right] {\scriptsize \textbf{$X_N$}};

                    \draw[thick] (-2,-1) circle (0.2);
                    \node at (-2,-1) {\textbf{\large $\times$}};

                    \draw[thick,->,mygreen] (0,1) -- (1,1) node[above, yshift = -3 mm] {\tiny \textbf{$Y_{L_{1}}$}};
                    \draw[thick,->,red] (0,1) -- (0,2) node[above, yshift= -3 mm] {\tiny \textbf{$X_{L_{1}}$}};
                    \draw[thick] (0,1) circle (0.1);
                    \node at (0,1) {\textbf{\small $\times$}};

                    \draw[thick,->,mygreen] (-1.5,0) -- (-1.5,1) node[right, xshift = -1 mm] {\tiny \textbf{$Y_{L_{2}}$}};
                    \draw[thick,->,red] (-1.5,0) -- (-0.5,0) node[above, yshift= -3 mm] {\tiny \textbf{$X_{L_{2}}$}};
                    \draw[thick] (-1.5,0) circle (0.1);
                    \node at (-1.5,0) {\textbf{\small $\times$}};

                \end{scope}
            \end{tikzpicture}    
          };
        
        \draw[thick] (0.45,0.5) -- ($(chatbox.west)$);
    \end{tikzpicture}
   \caption{Sample representation of global (ECEF), local (NED), and LiDAR sensor coordinate systems.}
    \label{fig: coord_frames}
    \end{figure}
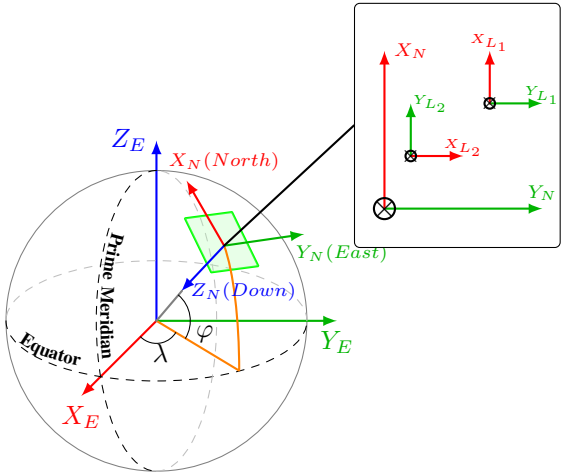

    \subsection{Zone-based TMC Estimation}

\begin{figure}[!hbt]
            \centering
            \includegraphics[width=0.9\linewidth]{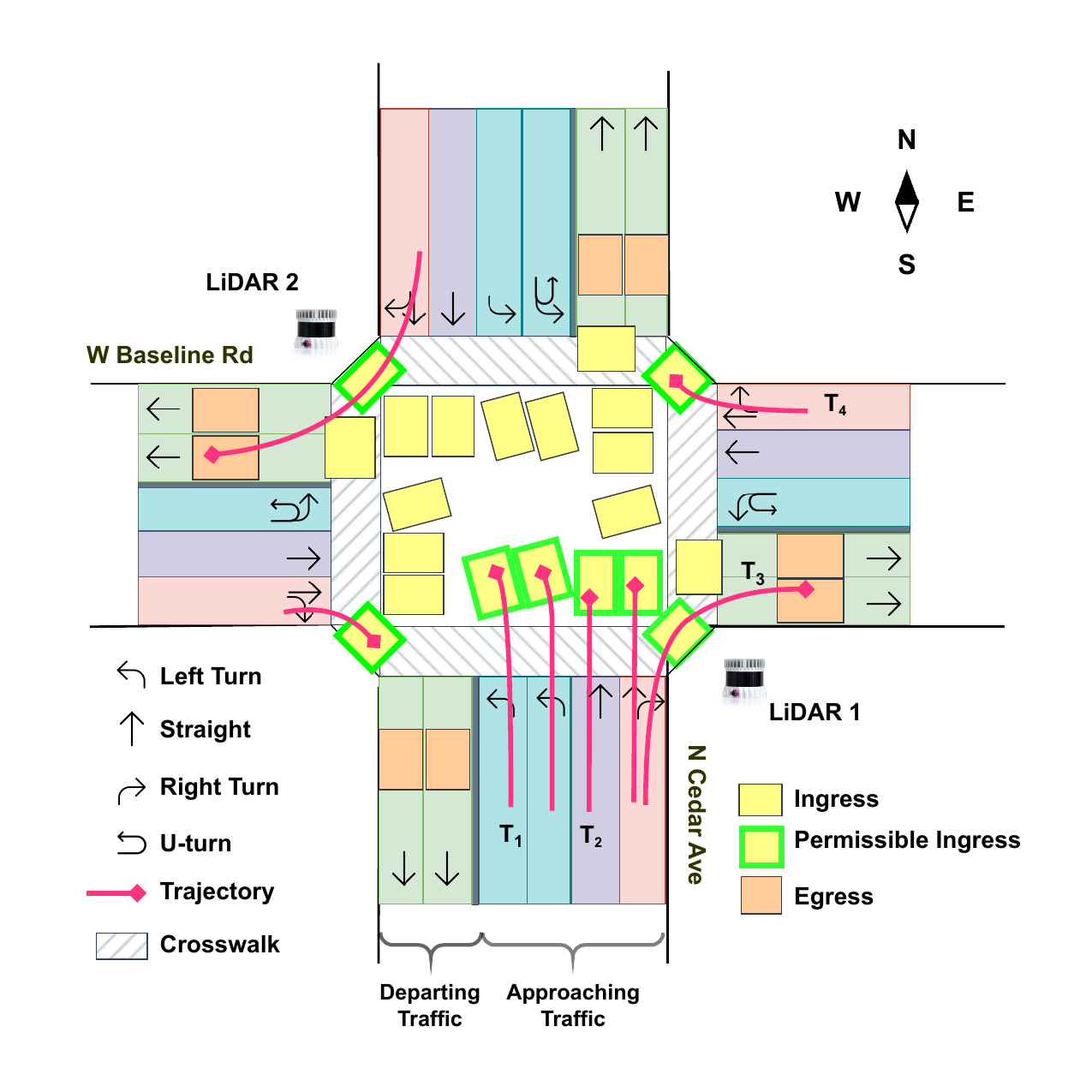}
            \caption{TMC calculation based on ingresses and egresses at the intersection.}
            \label{fig: zone_based_tmc}
        \end{figure}

        In zone-based TMC estimation, the vehicles are detected within a rectangular zone on the ingress and egress side of the traffic movements, as shown in Figure~\ref{fig: zone_based_tmc}. A zone is only triggered when a detection (bounding box centroid position) falls inside the zone, and the movement corresponding to the zone is permissible during a signal phase. The Signal Phase and Timing (SPaT) information is used to determine permissible zones. Right turns are allowed on red, and hence, the right-turn zones are permissible at all times. A specific signal phase involving only Northbound movements is illustrated in the Fig.~\ref{fig: zone_based_tmc}. 
        
        Vehicle counts for different movements (thru/left/u/right) are estimated from the time-series data representing detections in the ingress zones, e.g., $(T_1, T_2, T_4)$. No detections are observed at the northwest and southeast corners due to the limited field of view (FOV) of the LiDARs at those locations and the restricted detection range of the LiDARs positioned at the opposite corners. Due to this, the TMCs for northbound and southbound right turns are extracted from the egress zone time-series data $(T_3)$ as shown in Fig.~\ref{fig: zone_based_tmc}.

        LiDAR detections are generated at a frequency of 3–5 Hz, resulting in approximately 250 milliseconds between consecutive frames. Given a maximum free-flow speed of 20 meters per second (approximately 45 mph), the TMC detection zone was designed with a length of 8 meters. At lower speeds, a single vehicle may trigger the detection zone multiple times. However, based on empirical observations, the typical time headway between two vehicles is at least 1.5 seconds. Leveraging this insight, the minimum time thresholds of 2 seconds for right-turns and 1.2 seconds for other movements were applied to determine whether a subsequent trigger corresponds to a different vehicle. Additionally, detections occurring close to each other on the temporal scale were clustered and counted as a single vehicle to avoid overestimation of TMC.

\section{RESULTS AND DISCUSSIONS}
\label{sec: results_discussion}

    \subsection{Ground Truth (GT) Calculation}

        \begin{figure}[!hbt]
            \centering
            \includegraphics[width=0.6\linewidth]{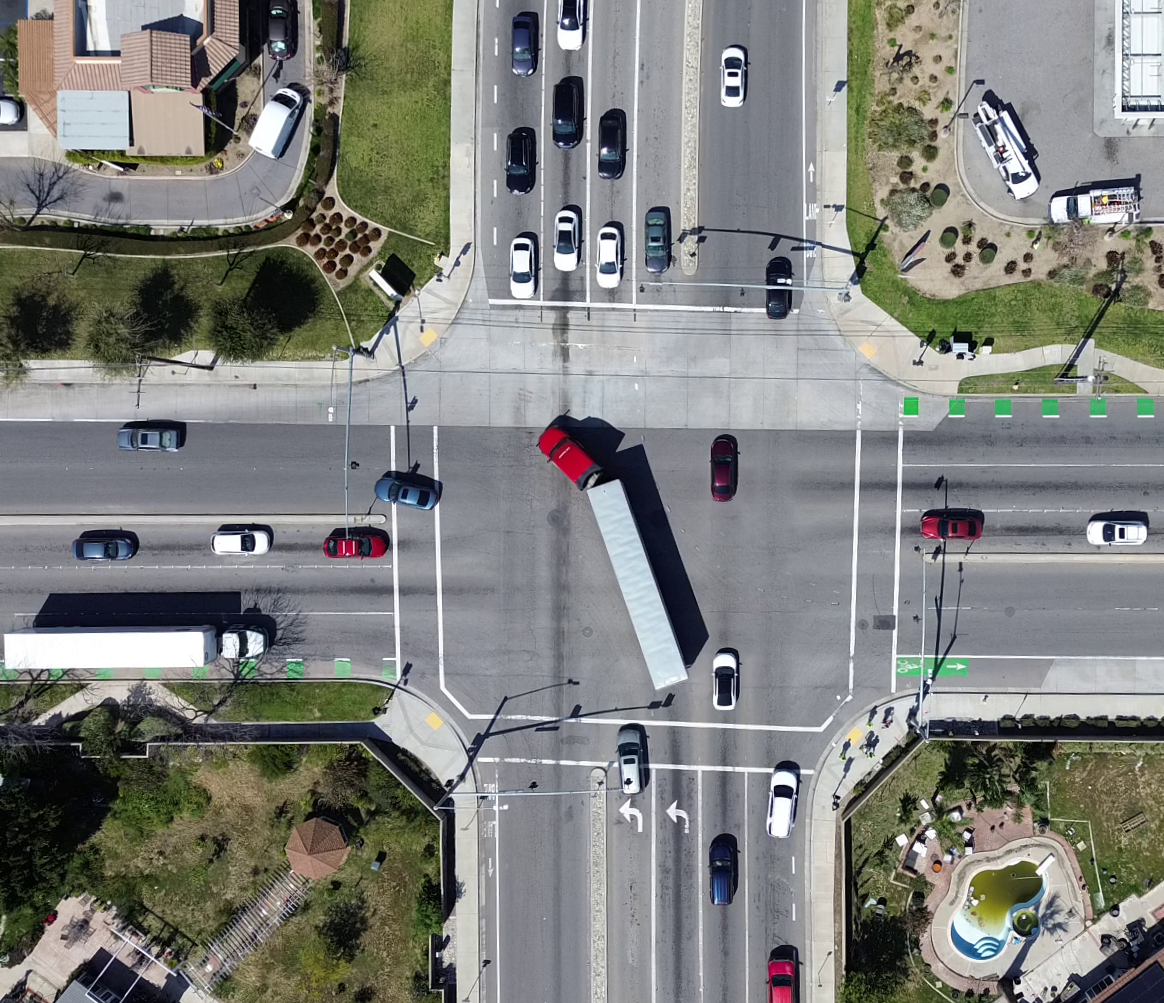}
            \caption{A snapshot of drone footage taken at the intersection of Baseline and Cedar.}
            \label{fig: drone_footage}
        \end{figure}
        
        To evaluate the accuracy of TMC estimation results, the LiDAR-estimated TMC is compared with the TMC calculated from the drone footage. The TMC calculation from the drone footage (in Fig.~\ref{fig: drone_footage}) is carried out through manual inspection. This calculated TMC shall be considered as the ground truth or the true TMC for analysis, assuming that human error in visual inspection is negligible. The true TMC is calculated for northbound (NB), southbound (SB), eastbound (EB) and westbound (WB) approaches for different movements, i.e., right $\&$ left turn, thru, U-turn, etc., and different vehicle types (class 1-6) as defined in TABLE~\ref{tab: vehicle_class}. 

        \begin{table}[h!]
            \centering
            \renewcommand{\arraystretch}{1}
            \caption{Vehicle Classification by Length and FHWA Classes}
            \label{tab: vehicle_class}
            \begin{tabular}{|p{0.5cm}|p{1cm}|p{3.5cm}|p{1.5cm}|}
                \hline
                \textbf{Class} & \textbf{Length (m)} & \textbf{Type} & \textbf{FHWA Class} \\
                \hline
                1 & $\leq 1$ & Pedestrians & N.A \\
                \hline
                2 & 1-2.2 & Bicycle, Scooterist, Motorbikes & Class 1 \\
                \hline
                3 & 2.2-5 & Hatchbacks, Sedans, Small-Medium SUVs & Class 2 (No Trailer) \\
                \hline
                4 & 5-7 & Large SUVs, Vans, Pickup Trucks & Class 2 (Trailer), Class 3 \\
                \hline
                5 & 7-12 & Trucks, Buses & Class 4, 5, 6, 7 \\
                \hline
                6 & $\geq 12$ & Trailers, Combination Trucks & Class 8, 9, 10 \\
                \hline
                \multicolumn{4}{|c|}{FHWA class 11, 12, and 13 are not present in the collected dataset} \\
                \hline
            \end{tabular}
        \end{table}

        \begin{table}[h!]
            \centering
            \caption{True TMC (Ground Truth) from Drone footage at the intersection of Baseline and Cedar}
            \label{tab: GT_TMC}
            \begin{tabular}{@{}p{0.7cm} p{0.7cm} *{4}{p{0.6cm}} p{0.9cm} p{0.6cm}@{}}
                \toprule
                \textbf{Time} & \textbf{Vehicle class} & \multicolumn{5}{c}{\textbf{Northbound}} & \textbf{...} \\
                \cmidrule(lr){3-7}
                \textit{(mins)} & \textit{(1-6)} & \textbf{Left} & \textbf{Thru} & \textbf{Right} & \textbf{U\--turn} & \textbf{Total} & \\
                \midrule
                \multirow{6}{*}{0 to 5} & 1 & 0 & 0 & 0 & 0 & 0 & ...\\
                & 2 & 0 & 0 & 0 & 0 & 0 & ...\\
                & 3 & 3 & 38 & 6 & 2 & 49 & ...\\
                & 4 & 1 & 11 & 2 & 1 & 15 & ...\\
                & 5 & 0 & 0 & 0 & 0 & 0 & ...\\
                & 6 & 0 & 0 & 0 & 0 & 0 & ...\\
        
                \midrule
                 ... & ... & ... & ... & ... & ... & ... & ... \\
                \bottomrule
            \end{tabular}
        \end{table}

        The data were collected from 1:51 pm to 2:10 pm (local time) on 20th March, 2024 and the TMC results are grouped into time intervals of 5 minutes as shown in TABLE~\ref{tab: GT_TMC}. It should be noted that in the ground truth TMC data, the vehicles were classified visually by the first author based on the type of vehicle, e.g., sedan, SUV, trailer, etc. Whereas, the vehicle class in the LiDAR-estimated TMC data is determined based on the length of the detected bounding box. As shown in TABLE~\ref{tab: GT_TMC}, detected objects are categorized into six groups.

    \subsection{Aggregated TMC results}

        \begin{figure*}[!ht]
            \centering
            \includegraphics[width=\textwidth]{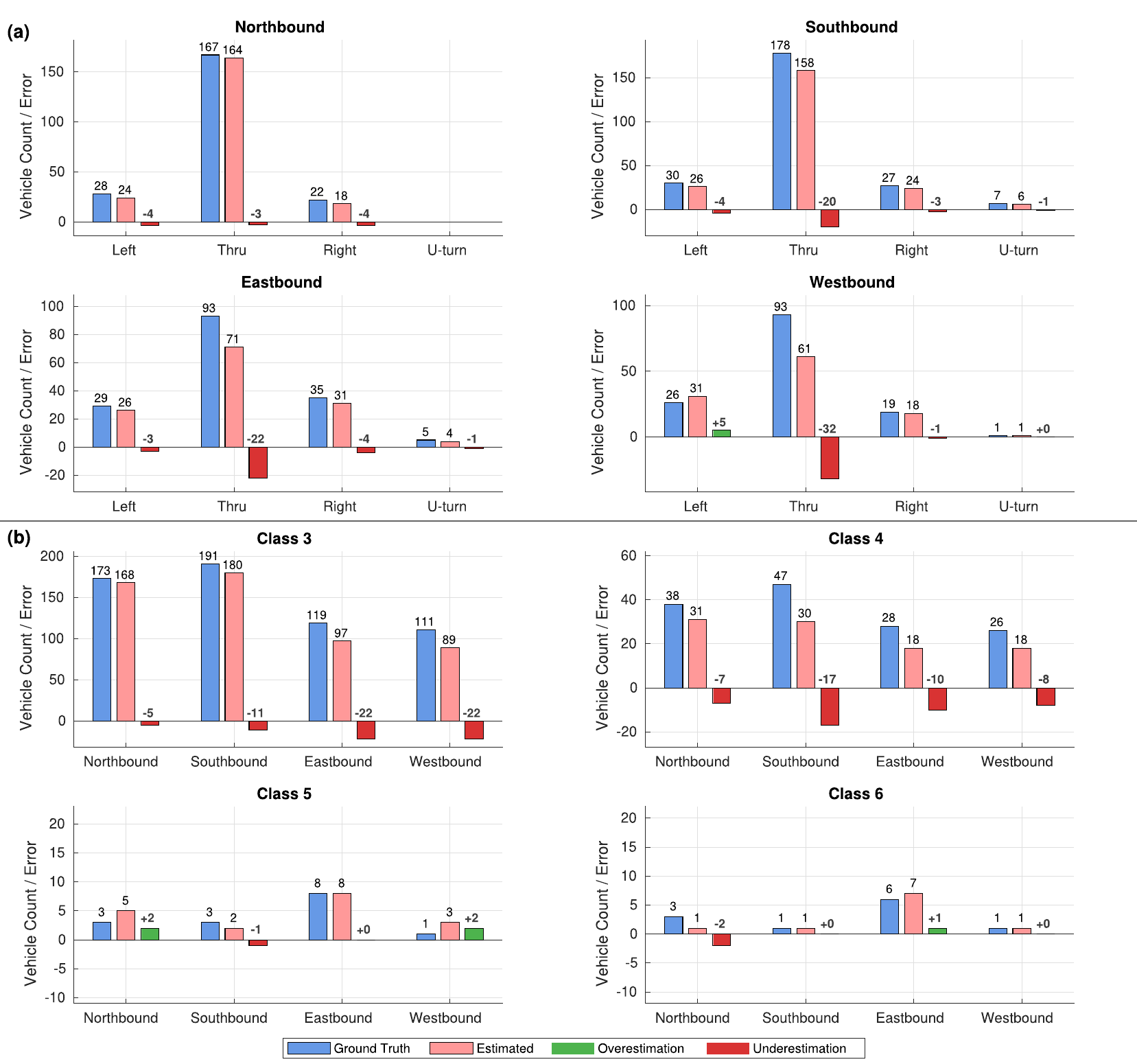}
            \caption{TMC comparison: (a) Travel Directions vs. Movements, (b) Vehicle Classes vs. Travel Directions}
            \label{fig:agg_tmc}
        \end{figure*}
        
        The TMC results are aggregated into two categories for analysis and visualization to simplify the analysis process. Firstly, the data are aggregated over time and across different vehicle classes to generate plots for vehicle counts and associated errors for different travel directions and movements. Similarly, the data are also aggregated over time and across different movements to generate plots for different vehicle classes and travel directions. All the results are depicted in Fig.~\ref{fig:agg_tmc}.

        Upon careful observation of the TMC plots, the following trends and irregularities were noted:
        
        \begin{enumerate} 
        
        \item Thru movement appears to be the dominant movement across various travel directions, with the majority of traffic moving in the north-south direction. 
        
        \item The eastbound and westbound thru-movement count estimation errors are relatively higher compared to the northbound and southbound thru-movement count errors. This is due to the farther placement of the dual LiDARs from the eastbound/westbound thru-ingress zones. Because of the greater distance and occasional occlusions by vehicles in adjacent lanes, the LiDAR-based detection algorithm could not consistently generate bounding boxes in these ingress zones. 
        
        \item Upon analyzing the TMC for different vehicle classes across various travel directions, it was found that vehicle class 3 (81.03\%) and vehicle class 4 (14.72\%) constitute the majority of traffic. 
        
        \item The TMC error for class 4 vehicles (30.21\% of the total vehicle volume) is higher than that for class 3 vehicles (10.1\%). The relative increase in TMC error for class 4 vehicles is attributed to vehicle classification errors during ground truth calculations. Vehicles such as medium- and large-sized SUVs, whose bounding box lengths were greater than 5 meters, were incorrectly labeled as class 3 vehicles by the human annotator based on drone footage visuals. 
        
        \item The TMC errors for class 5 and class 6 vehicles are smaller. The TMC is overestimated for certain groups (class 5 – northbound, westbound; class 6 – eastbound). This overestimation arises from the fact that heavier vehicles take longer to cross the ingress detection zone, leading to additional counts within the specified time threshold.
       
        \end{enumerate}

    Based on the results, it is inferred that the estimated TMC can be used to analyze traffic movement counts, flow, volume, etc. However, the TMC is affected by errors such as LiDAR-based detection errors, fixed time thresholds for detecting vehicles in ingress zones, and vehicle misclassification. These errors will be defined, categorized, and quantified for detailed analysis in future work.

\section{LESSONS LEARNED AND POTENTIAL IMPROVEMENTS}
\label{sec: lessons_learned}

The estimated TMC results using LiDARs are satisfactory; however, several limitations need to be addressed for future improvements.

\begin{enumerate}
    \item The purpose of using dual LiDARs was to cover the entire intersection. However, the detection range is limited to approximately 40 meters for the specific LiDAR (OS1-128, Rev.6) and the detection algorithm in use, i.e., PointPillars. Due to this limitation, it was not possible to detect vehicles at the corners where the LiDARs were installed. It is advisable to experiment with different LiDAR configurations (placements, resolution, etc.) and state-of-the-art algorithms to address this issue.
    
    \item In this study, only detection parameters such as position and length were used for TMC estimation. While this method is simple and easy to implement, it requires careful definition and fine-tuning of ingress zones and vehicle counting criteria. More sophisticated methods, such as tracking-by-detection, could be employed to track individual road agents and classify their movements for improved TMC estimation. This would help in tapping the full potential of the LiDARs.
    
    \item Collecting raw point cloud data and running the detection pipeline simultaneously for two LiDARs resulted in a reduction in the data storage frequency. Although the original point cloud data is generated at 10~Hz, the processed raw dataframes and detections were stored at 3--5~Hz for each LiDAR due to the overall detection pipeline processing time. This could be improved by implementing multi-threading techniques to handle multiple detection pipelines for multiple sensors concurrently.
    
    \item End-to-end detection, tracking, and intent prediction algorithms could be explored to track vehicles effectively with low (point cloud data) frame rates. Furthermore, end-to-end models could offer improved feature-level data association, potentially resolving data association issues encountered with classical tracking-by-detection approaches.
    
\end{enumerate}

Apart from the noted limitations, the use of LiDARs for traffic surveillance and estimation remains highly attractive. Various traffic states, ranging from macroscopic to microscopic levels, can be accurately estimated using LiDARs. Moreover, due to their inherent operating principle, LiDARs address many privacy concerns typically associated with other surveillance technologies, such as cameras. LiDARs are also resilient to changing weather and lighting conditions, making them an effective and reliable alternative for traffic monitoring applications.

\section{CONCLUSIONS}
\label{sec: conclusion}

This paper presented the deployment of a dual-LiDAR system and provided a preliminary analysis of the estimated Traffic Movement Count (TMC) at a signalized intersection. The detected bounding box positions and dimensions from the PointPillars model were used to identify vehicles passing through predefined ingress and egress zones. The resulting vehicle count time series data was then analyzed to estimate TMC across different vehicle classes and movements. The analysis also revealed that the TMC estimation process is affected by various sources of error, including the limited range of the LiDARs, missed detections, vehicle misclassification, and fixed thresholds in the zone-based counting method. To fully leverage the capabilities of LiDAR technology, advanced methods such as tracking-by-detection and end-to-end tracking with intent prediction should be explored in future work. Additionally, several lessons learned from the field deployment and analysis were presented, which may benefit researchers and practitioners undertaking similar experiments.

\section*{ACKNOWLEDGEMENT}
This material is based upon work supported by the Southern California Association of Governments, under Agreement Number 22-025-C01. The authors acknowledge the support received from the staff of the City of Rialto during the field experiment.


\bibliographystyle{IEEEtran}
\bibliography{references.bib}

\end{document}